\documentclass[10pt,aps,prx,twocolumn,superscriptaddress,floatfix]{revtex4-2}
\usepackage{braket}
\usepackage{amsfonts, bm, mathtools}
\usepackage{booktabs}
\usepackage{siunitx}
\usepackage{xcolor}
\usepackage{xurl}
\usepackage[breaklinks=true,colorlinks,citecolor=blue,linkcolor=blue,urlcolor=blue]{hyperref}
\usepackage{txfonts}
\usepackage{orcidlink}
\usepackage{float}
\usepackage{algorithm}
\usepackage{algorithmic}



\begin{document}

\title{Velocity Verlet-Based Optimization for Variational Quantum Eigensolvers}

\author{Rinka Miura$^{\orcidlink{0009-0003-6502-2243}}$}
\email[]{acdccit5@gmail.com}
\affiliation{Department of Applied Chemistry, Kobe City College of Technology, 8 Chome-3 Gakuen Higashimachi, Nishi Ward, Kobe, 651-2102, Hyogo, Japan}

\date{\today}

\begin{abstract}
The Variational Quantum Eigensolver (VQE) is a key algorithm for near-term quantum computers, yet its performance is often limited by the classical optimization of circuit parameters. We propose using the velocity Verlet algorithm, inspired by classical molecular dynamics, to address this challenge. By introducing an inertial "velocity" term, our method efficiently explores complex energy landscapes. We compare its performance against standard optimizers on H$_2$ and LiH molecules. For H$_2$, our method achieves chemical accuracy with fewer quantum circuit evaluations than L-BFGS-B. For LiH, it attains the lowest final energy, demonstrating its potential for high-accuracy VQE simulations.
\end{abstract}

\maketitle

\section{Introduction}

Hybrid quantum-classical variational algorithms have emerged as a promising path for quantum chemistry on near-term devices, with the Variational Quantum Eigensolver (VQE) providing a flexible framework to approximate ground states using parametrized quantum circuits optimized by classical routines~\cite{peruzzo2014, mcclean2016}. Nevertheless, VQE performance is limited by nonconvex and often rugged energy landscapes, the phenomenon of barren plateaus where gradients vanish exponentially with system size~\cite{mcclean2018, cerezo2021cost, holmes2021}, and the substantial measurement overhead associated with Hamiltonian expectation estimation~\cite{zoltan2025}. Recent reviews provide broad surveys of algorithmic strategies and resource considerations for quantum computational chemistry~\cite{cao2019, bharti2022, cerezo2021vqa, endo2021, tilly2022}.

A considerable body of work has been devoted to improving VQE optimization. Geometry-aware updates, including the quantum natural gradient rooted in the Fubini–Study metric, can align descent with the underlying quantum-state manifold~\cite{stokes2020}. Quasi-Newton methods such as L-BFGS-B are widely used because of their rapid local convergence and moderate memory footprint~\cite{zhu1997}, while derivative-free optimizers like COBYLA can be favorable on hardware runs as they avoid costly gradient estimation~\cite{powell1994, berahas2019}. Analytic single-parameter update schemes, exemplified by sequential minimal optimization and related “rotosolve”-style updates, exploit the sinusoidal dependence of certain gates to produce closed-form optima for one parameter at a time~\cite{schuld2019, mitarai2018}. Complementary to advances in optimization, many works seek to reduce measurement costs using commutativity-based grouping, classical shadows, and other statistical and algebraic methods~\cite{zoltan2025, krovi2022}. Meanwhile, several approaches aim to move beyond standard outer-loop VQE by incorporating subspace methods for excited states and error mitigation~\cite{mcclean2017}, and time-evolution-inspired Krylov or Lanczos subspace techniques~\cite{stair2020, sugisaki2025, mikkelsen2025, yu2025}, as well as imaginary-time evolution and variational simulation dynamics that trade optimization for physics-informed updates~\cite{motta2020, yuan2019, mcardle2019}.

Inspired by algorithms from classical molecular dynamics, the present work connects the landscape-based interpretation of molecular optimization~\cite{wales2004, nocedal2006} with recent advances in quantum-classical hybrid algorithms~\cite{ostaszewski2021, nakanishi2019, benedetti2019, obrien2019, zhang2020}, positioning the velocity Verlet optimizer within the broader context of quantum machine learning in feature Hilbert spaces~\cite{schuld2019qml}. The velocity Verlet integrator is a second-order accurate scheme with excellent long-time stability for Hamiltonian systems~\cite{verlet1967}, foundational in simulating atomistic dynamics~\cite{hairer2006}. Here, we reinterpret VQE parameter vectors as generalized positions, introduce velocities as auxiliary variables, and define forces as the negative energy gradient. To promote convergence in an optimization setting rather than preserve energy as in physical simulation, we include linear damping to dissipate kinetic energy. The resulting update inherits an inertial mechanism akin to heavy-ball momentum~\cite{polyak1964}, potentially helping to traverse shallow local traps and damp oscillations in narrow valleys, while deliberately sacrificing time-reversibility and symplecticity to ensure dissipation.

To characterize the performance profile and identify the potential utility of this physics-inspired approach, we evaluate the proposed method on two molecular benchmarks at equilibrium geometries: H$_2$ in the STO-3G basis mapped to 4 qubits and LiH in the STO-3G basis mapped to 12 qubits. The ansatz is hardware-efficient with four alternating layers of single-qubit rotations and CZ entanglers~\cite{kandala2017}. We compare against L-BFGS-B, SLSQP, and COBYLA using exact statevector simulations. We adopt the number of energy evaluations as a principal performance metric, since it is a proxy for quantum resource cost and dominates runtime in real-device experiments~\cite{zoltan2025}. For H$_2$, the Verlet-based optimizer reaches chemical accuracy with fewer evaluations than L-BFGS-B within the plotted 40-iteration budget, and it attains the smallest final error among all tested methods under that budget. For LiH, none of the optimizers achieve chemical accuracy within the same iteration budget, but the proposed method yields the lowest final error. We provide a careful discussion of hyperparameter sensitivity, the role of damping and inertia, and the trade-offs associated with gradient estimation. We conclude with limitations and possible improvements, including single-force variants to reduce evaluation counts, adaptive hyperparameter schemes, and an outlook toward experiments on noisy hardware.

\section{Theory and Methods}

\subsection{Variational Quantum Eigensolver}
Let $H$ denote the electronic Hamiltonian mapped to qubits. A parametrized unitary $U(\theta)$ acting on the reference $\ket{0}$ prepares a trial state $\ket{\psi(\theta)} = U(\theta)\ket{0}$ with energy expectation $E(\theta) = \braket{\psi(\theta) | H | \psi(\theta)}$. By the variational principle, $E(\theta) \geq E_0$ for all $\theta$, where $E_0$ is the ground-state energy. VQE seeks $\theta^* = \arg\min_\theta E(\theta)$, using a classical optimizer in an outer loop to update $\theta$ based on energy and, when available, gradient information~\cite{peruzzo2014, mcclean2016, cao2019, bharti2022, cerezo2021vqa, endo2021, tilly2022}.

\subsection{Velocity Verlet optimization with damping}
We draw an analogy to classical dynamics by identifying the parameter vector $\theta$ as a generalized position, introducing an auxiliary velocity $v$ of the same dimension, and defining the force as $F(\theta) = -\nabla E(\theta)$. In the undamped Velocity Verlet scheme for a particle of mass $m$ and timestep $\Delta t$, the updates read
\begin{align*}
v(t+\Delta t/2) &= v(t) + \Delta t F(\theta(t))/(2m), \\
\theta(t+\Delta t) &= \theta(t) + \Delta t v(t+\Delta t/2), \\
v(t+\Delta t) &= v(t+\Delta t/2) + \Delta t F(\theta(t+\Delta t))/(2m).
\end{align*}
This scheme is time-reversible and symplectic for conservative forces~\cite{verlet1967, hairer2006}. In optimization, however, one seeks to dissipate rather than conserve the ``kinetic energy'' associated with $v$. We therefore introduce linear damping with coefficient $\gamma \in[0,1)$, which produces a contractive map on $v$ and breaks both time-reversibility and symplecticity. In geometric numerical integration terms, such damping leads to a conformally symplectic, dissipative integrator rather than a symplectic one~\cite{hairer2006}. Dissipation is desirable here because it prevents persistent oscillations and allows the trajectory to settle into a local minimum.

In our implementation, damping is applied to the post-update velocity, and the forces are constructed from numerical gradients as described below. The velocity imparts inertia that can smooth high-curvature directions and help traverse shallow basins, a mechanism related to classical heavy-ball momentum~\cite{polyak1964}, while the damping term suppresses overshoot and ringing, especially near narrow valleys.

\subsection{Molecular systems, mappings, and exact references}
We considered H$_2$ at an internuclear distance of 0.977 \AA~and LiH at 1.596 \AA, each in the STO-3G basis. Molecular integrals and full configuration interaction (FCI) reference energies were obtained with PySCF~\cite{sun2018} through the OpenFermion–PySCF interface~\cite{mcclean2020}. The second-quantized Hamiltonians were mapped to qubit operators via the Jordan–Wigner transform~\cite{jordan1928} as implemented in OpenFermion~\cite{mcclean2020} and converted to Qulacs observables~\cite{suzuki2021}. The resulting qubit counts and electron numbers were 4 qubits and 2 electrons for H$_2$, and 12 qubits and 4 electrons for LiH. The exact ground-state energies used for error computation were $-1.1059333523$ Hartree for H$_2$ and $-7.8823869936$ Hartree for LiH, as reported by PySCF at these geometries. All expectation values were computed via noiseless, exact statevector simulation using Qulacs~\cite{suzuki2021}, without measurement shot noise.

\subsection{Ansatz and initialization}
We employed a hardware-efficient ansatz of depth four~\cite{kandala2017}. Each layer comprised $R_Y$ and $R_Z$ rotations on every qubit followed by a pattern of CZ entanglers arranged in a ladder across neighboring qubits, with a final rotation layer appended. With this structure, the number of real parameters was $2n(\text{depth}+1)$, namely 40 for H$_2$ and 120 for LiH. The initial parameters were drawn i.i.d. from a uniform distribution on $[0,0.1)$ using a fixed seed to ensure identical initialization across all optimizers. The seed was set to 18.

\subsection{Gradient Estimation and Evaluation-Count Accounting}
The energy gradients were computed using the parameter-shift rule, which enables exact analytical differentiation on quantum hardware without resorting to finite-difference approximations. This approach, initially introduced by~\cite{schuld2019} and~\cite{mitarai2018}, provides unbiased estimates of the expectation-value derivatives by evaluating the circuit at shifted parameters. Consequently, the total number of circuit evaluations scales linearly with the number of variational parameters~\cite{schuld2019, mitarai2018}. Consequently, the total number of circuit evaluations scales linearly with the number of variational parameters. The gradient-based optimization process and its associated evaluation cost are outlined in Algorithm 1.

\begin{algorithm}[H]
\caption{Gradient-based VQE optimization using parameter-shift rule}
\begin{algorithmic}[1]
\STATE \textbf{Input:} Hamiltonian $H$, Ansatz circuit $U(\theta)$, initial parameters $\theta_0$, optimizer $\mathcal{O}$
\STATE \textbf{Initialize} $\theta \leftarrow \theta_0$
\REPEAT
    \STATE // Energy evaluation
    \STATE $E(\theta) \leftarrow \braket{0 | U^\dagger(\theta) H U(\theta) | 0}$
    \STATE // Gradient evaluation (parameter-shift rule)
    \FOR{each parameter $\theta_i$}
        \STATE $E^+ \leftarrow \braket{0 | U^\dagger(\theta \text{ with } \theta_i \to \theta_i + \pi/2) H U(\theta \text{ with } \theta_i \to \theta_i + \pi/2) | 0}$
        \STATE $E^- \leftarrow \braket{0 | U^\dagger(\theta \text{ with } \theta_i \to \theta_i - \pi/2) H U(\theta \text{ with } \theta_i \to \theta_i - \pi/2) | 0}$
        \STATE $\partial E / \partial \theta_i \leftarrow (E^+ - E^-) / 2$
    \ENDFOR
    \STATE $\theta \leftarrow \mathcal{O}(\theta, \nabla E(\theta))$ // classical update step
\UNTIL{convergence}
\STATE \textbf{Output:} Optimized parameters $\theta^*$
\end{algorithmic}
\end{algorithm}

\noindent Total number of circuit evaluations per iteration:
\begin{equation}
N_{\text{eval\_per\_iter}} = 1 \text{ (for energy)} + 2 \times N_{\text{param}} \text{ (for gradients)}
\end{equation}

We adopt ``number of energy evaluations'' as a principal performance metric and explicitly count every call to the energy function. For COBYLA, which is gradient-free, only function calls were counted~\cite{powell1994, berahas2019}. Where we report ``Total Evals'' in the performance tables below, we refer to the cumulative energy evaluations accrued over the 40-iteration window plotted in our convergence figures. Note that on actual quantum hardware, a single energy evaluation requires measuring multiple Pauli terms derived from the Hamiltonian via the Jordan–Wigner transform, implying an additional measurement overhead.

\subsection{Optimizers, hyperparameters, and stopping conditions}
The velocity Verlet update used the following hyperparameters determined by preliminary tuning: for H$_2$, $\Delta t=0.01$, $m=0.8$, $\gamma=0.68$; for LiH, $\Delta t=0.01$, $m=1.9$, $\gamma=0.68$. The initial velocity was set to zero. We compared against L-BFGS-B and SLSQP as implemented in SciPy~\cite{zhu1997, virtanen2020} with gradients supplied by the parameter-shift rule, and against COBYLA as a gradient-free baseline~\cite{powell1994, virtanen2020}. The outer-loop iteration budget for the plots and tables was set to 40 iterations for all methods. Each optimizer was run with identical initial parameters. All runs were executed on a single CPU, AMD EPYC 7B12, and the wall-clock times reported include all function and gradient calls. Chemical accuracy was defined as $1.6 \times 10^{-3}$ Hartree.

\section{Results}

\subsection{Hydrogen molecule (H$_2$, 4 qubits)}
\begin{figure}[!tb]
    \centering
    \includegraphics[width=\linewidth]{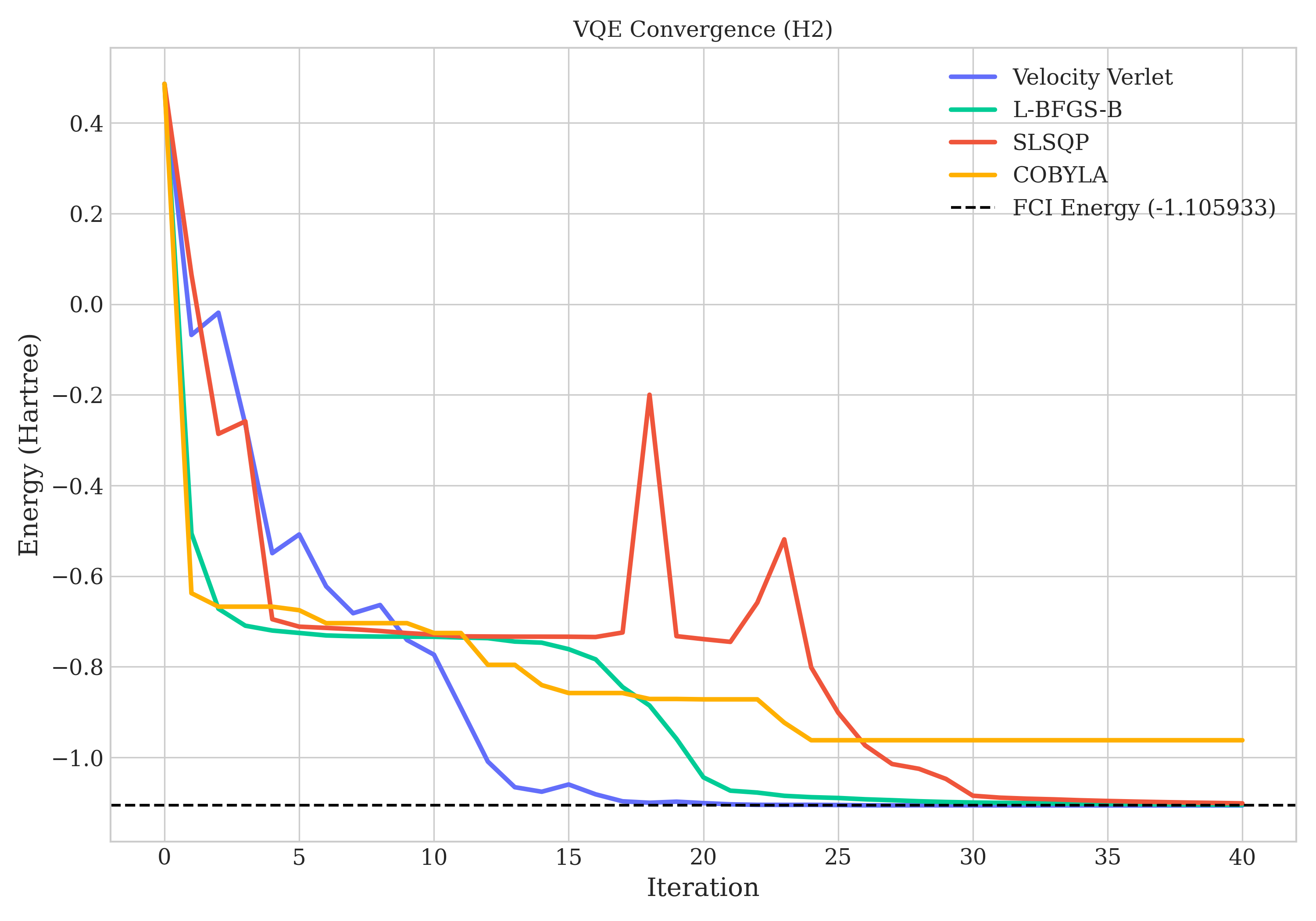}
    \caption{\textbf{Energy convergence as a function of optimization iteration for the H$_2$ molecule.} The plot displays the energy expectation value for the velocity Verlet, COBYLA, L-BFGS-B, and SLSQP optimizers. The dashed line represents the exact FCI energy.}
    \label{fig:1}
\end{figure}

The performance of the velocity Verlet optimizer was first evaluated on the hydrogen molecule. Figure~\ref{fig:1} illustrates the convergence of the energy expectation value as a function of the optimization iteration for each of the four tested methods. The horizontal axis represents the iteration number, while the vertical axis shows the calculated energy in Hartree. The dashed black line indicates the reference Full Configuration Interaction (FCI) energy ($-1.1059333523$ Hartree), which is the exact ground-state energy within the STO-3G basis set. As depicted, all optimizers successfully decrease the energy over the 40-iteration budget. Notably, the velocity Verlet method (blue line) exhibits a particularly rapid initial descent and converges smoothly to a final energy value that is visually indistinguishable from the FCI energy. In contrast, L-BFGS-B, SLSQP, and COBYLA show a more gradual convergence.

\begin{figure}[!tb]
    \centering
    \includegraphics[width=\linewidth]{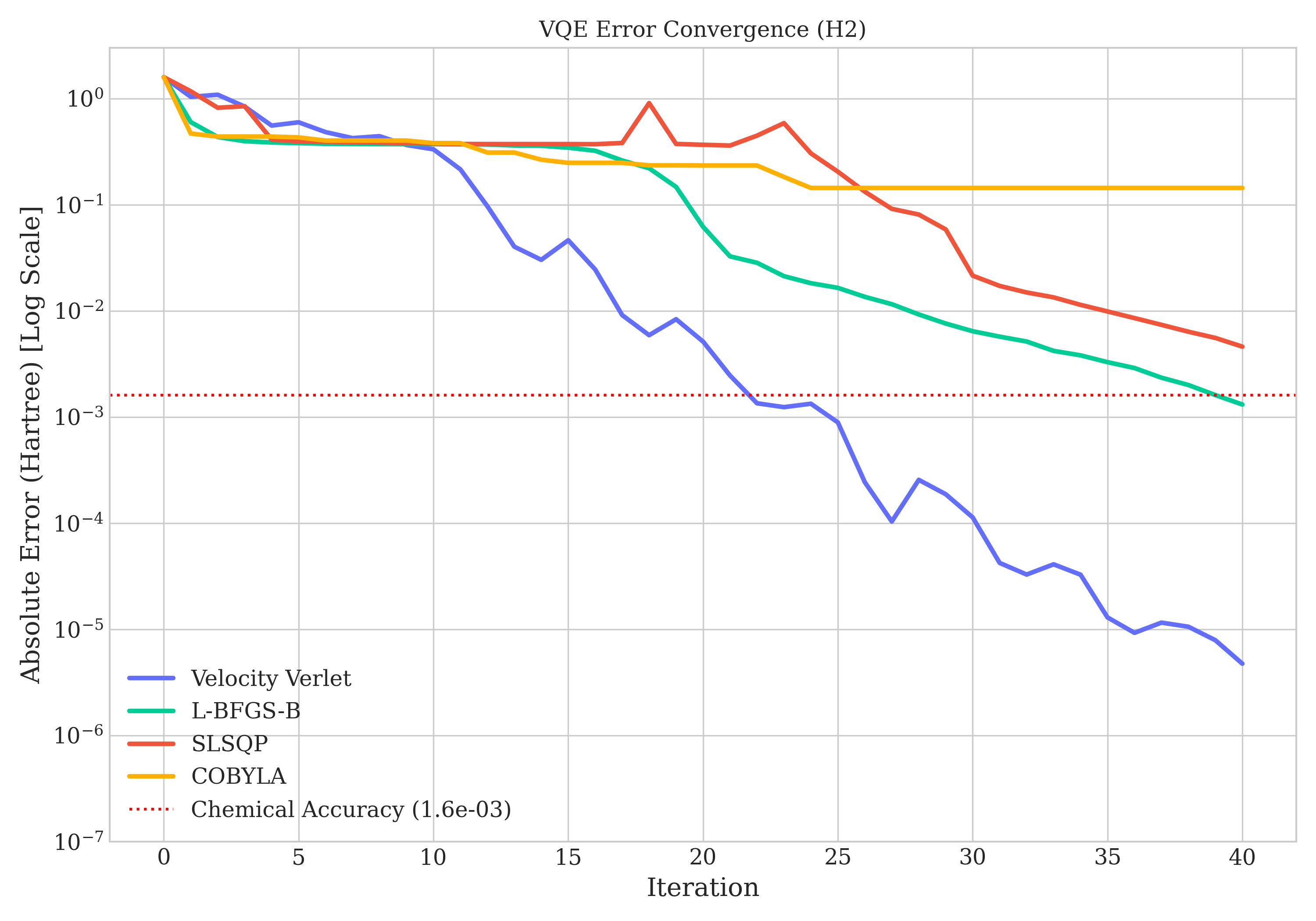}
    \caption{\textbf{Absolute error versus the number of energy evaluations for the H$_2$ molecule.} The vertical axis is on a logarithmic scale. The velocity Verlet method achieves chemical accuracy (red dotted line) with a lower number of quantum circuit evaluations than L-BFGS-B.}
    \label{fig:2}
\end{figure}

To provide a more quantitative analysis of convergence in a metric that reflects quantum resource cost, Figure~\ref{fig:2} plots the absolute error, $\lvert E(\theta) - E_{\text{FCI}} \rvert$, as a function of the number of energy evaluations. The vertical axis is presented on a logarithmic scale to highlight differences in precision. This figure reveals that the velocity Verlet optimizer not only converges to a lower final error but also reaches the threshold for chemical accuracy ($1.6 \times 10^{-3}$ Hartree, indicated by the dotted red line) with fewer energy evaluations (3,543) than L-BFGS-B (4,253). This result underscores the potential cost-efficiency of the inertial approach for achieving high-precision solutions.

The key performance metrics for the H$_2$ simulation are compiled in Table~\ref{tab:1}. The data confirm the graphical observations. The velocity Verlet method achieves the lowest final error ($4.74 \times 10^{-6}$ Hartree), demonstrating superior accuracy. The table also quantifies the number of evaluations required to reach chemical accuracy, confirming the advantage of the velocity Verlet method in this regard. While the gradient-free COBYLA optimizer has the lowest number of evaluations within the 40-iteration window, its final error is the largest, indicating that it would require a significantly longer run to achieve comparable accuracy.

\begin{table*}[!tb]
    \caption{\textbf{Performance summary for H$_2$ (4 qubits).} N/A indicates that chemical accuracy was not reached within the 40-iteration limit.$^\dagger$}
    \label{tab:1}
    \begin{ruledtabular}
    \begin{tabular}{lccccc}
    \multicolumn{1}{c}{Optimizer} & \multicolumn{1}{c}{Final Energy (Ha)} & \multicolumn{1}{c}{Final Error (Ha)} & \multicolumn{1}{c}{Total Evals} & \multicolumn{1}{c}{Time (s)} & \multicolumn{1}{c}{Evals to Chem. Acc.} \\ \midrule
    Velocity Verlet & -1.1059286124 & 4.7399e-06 & 6441 & 1.09 & 3543 \\
    L-BFGS-B & -1.1046226121 & 1.3107e-03 & 4253 & 1.74 & 4253 \\
    SLSQP & -1.1013332227 & 4.6001e-03 & 3294 & 1.75 & N/A \\
    COBYLA & -0.9620675939 & 1.4387e-01 & 156 & 3.49 & N/A \\
    \end{tabular}
    \end{ruledtabular}
    \vspace{1ex}
    \raggedright \small
    $^\dagger$ The total number of circuit evaluations for gradient-based methods was estimated as $N_{\text{eval}} \approx N_{\text{iter}} (1+2N_{\text{param}})$, accounting for both energy and gradient measurements under the parameter-shift rule. The reported values are the exact counts from the simulation.
\end{table*}

\subsection{Lithium hydride (LiH, 12 qubits)}
\begin{figure}[!tb]
    \centering
    \includegraphics[width=\linewidth]{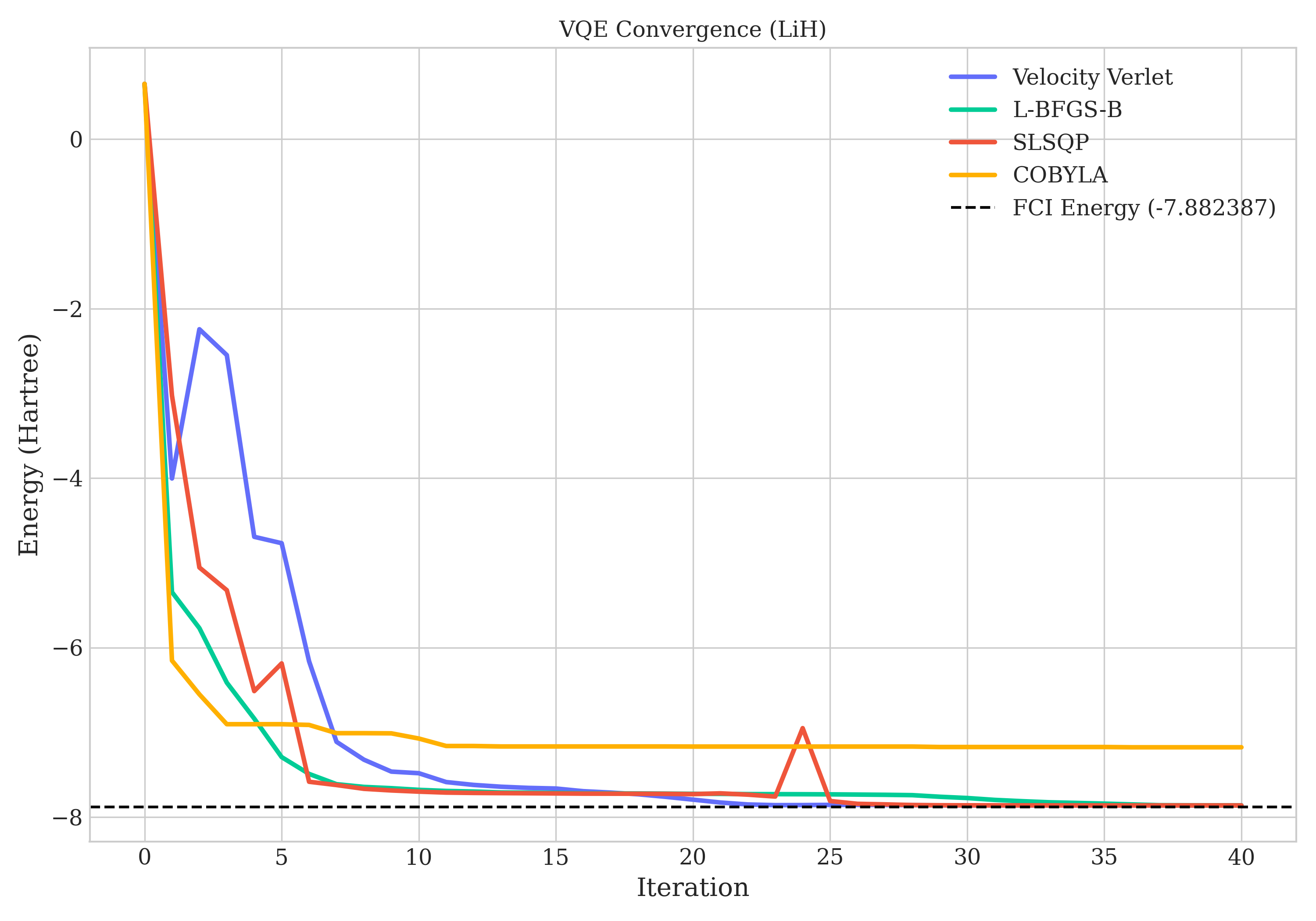}
    \caption{\textbf{Energy convergence as a function of iteration for the LiH molecule.} In this more complex landscape, the velocity Verlet method finds a substantially lower energy solution compared to the other optimizers.}
    \label{fig:3}
\end{figure}

To assess the scalability and performance on a more challenging problem, we next applied the optimizers to the 12-qubit LiH molecule. The energy convergence as a function of iteration is shown in Figure~\ref{fig:3}. The higher dimensionality and more complex energy landscape of this system make optimization more difficult for all methods. Within the 40-iteration limit, none of the optimizers successfully reached chemical accuracy. However, the velocity Verlet optimizer consistently converges to a lower energy state than L-BFGS-B, SLSQP, and COBYLA, ultimately achieving the lowest final error among the tested methods.

\begin{figure}[!tb]
    \centering
    \includegraphics[width=\linewidth]{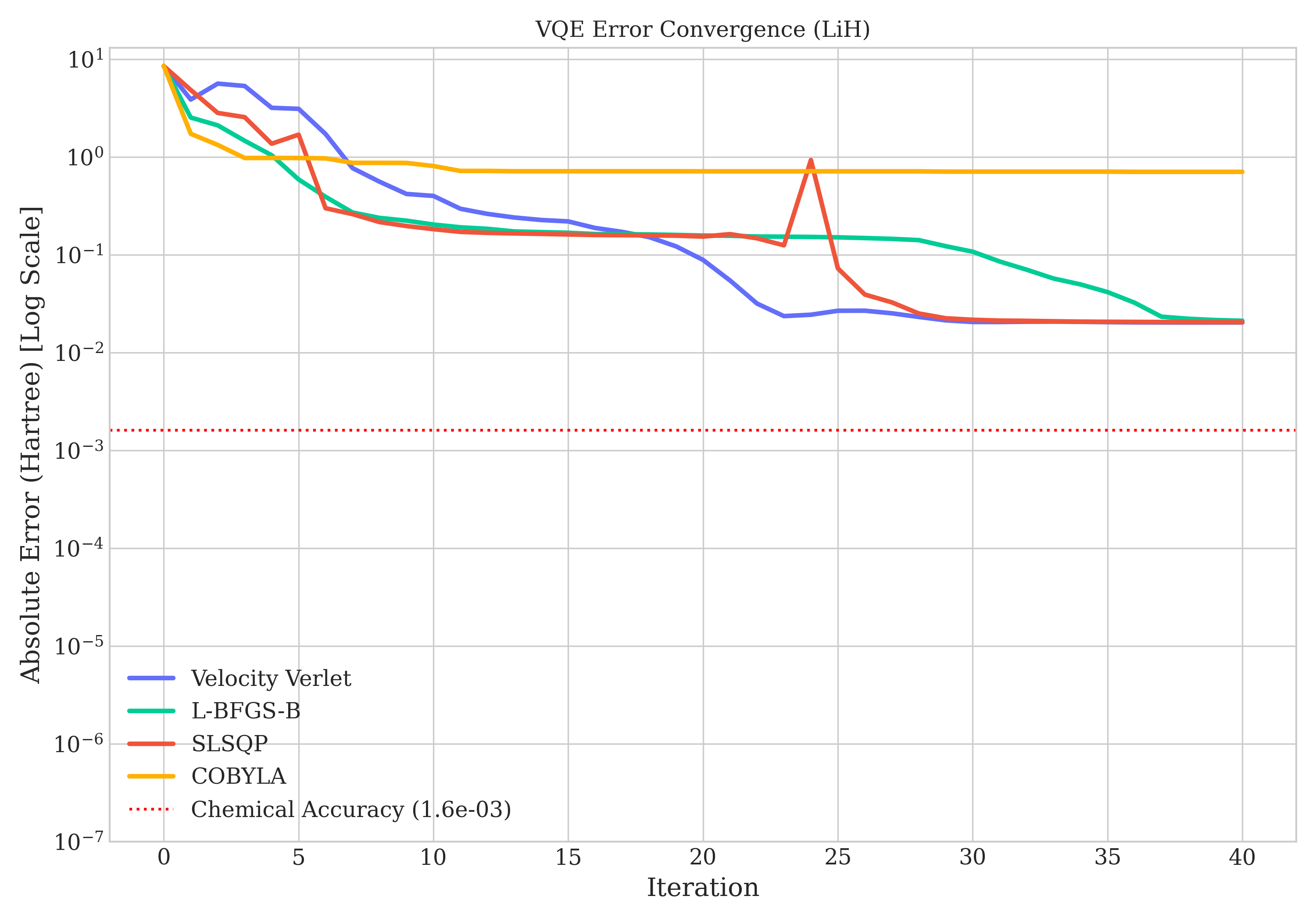}
    \caption{\textbf{Absolute error versus the number of energy evaluations for the LiH molecule.} Although no method reaches chemical accuracy, the velocity Verlet method consistently maintains the lowest error throughout the optimization.}
    \label{fig:4}
\end{figure}

The absolute error as a function of energy evaluations for LiH is presented in Figure~\ref{fig:4}. This plot further illustrates the advantage of the velocity Verlet method in terms of final accuracy, as it maintains the lowest error throughout the latter half of the optimization run.

Table~\ref{tab:2} provides a quantitative summary of the LiH results. The velocity Verlet method’s final error of $2.03 \times 10^{-2}$ Hartree is the best among the four optimizers. This improved accuracy comes at the cost of a higher number of total energy evaluations and, consequently, a longer wall-clock time. This highlights a critical trade-off, which is explored further in the Discussion section.

\begin{table*}[!tb]
    \caption{\textbf{Performance summary for LiH (12 qubits).} See Table~\ref{tab:1} for footnote definitions.}
    \label{tab:2}
    \begin{ruledtabular}
    \begin{tabular}{lccccc}
    \multicolumn{1}{c}{Optimizer} & \multicolumn{1}{c}{Final Energy (Ha)} & \multicolumn{1}{c}{Final Error (Ha)} & \multicolumn{1}{c}{Total Evals} & \multicolumn{1}{c}{Time (s)} & \multicolumn{1}{c}{Evals to Chem. Acc.} \\ \midrule
    Velocity Verlet & -7.8620500646 & 2.0337e-02 & 19241 & 254.10 & N/A \\
    L-BFGS-B & -7.8612173560 & 2.1170e-02 & 11127 & 148.80 & N/A \\
    SLSQP & -7.8619073396 & 2.0480e-02 & 9687 & 132.67 & N/A \\
    COBYLA & -7.1759086855 & 7.0648e-01 & 236 & 35.37 & N/A \\
    \end{tabular}
    \end{ruledtabular}
\end{table*}

\section{Discussion}

\subsection{Relation to the Variational Principle}
The velocity Verlet method, as adapted for VQE, remains entirely consistent with the variational principle. Each update to the parameter vector $\theta$ generates a new trial state, and its corresponding energy expectation value $E(\theta)$ is always guaranteed to be an upper bound to the true ground state energy $E_0$. The algorithm does not alter the objective function itself but rather provides a novel strategy for navigating its landscape. In essence, the velocity Verlet optimizer performs a momentum-assisted guided search within the parameter space, with the explicit goal of minimizing $E(\theta)$ and thereby approaching $E_0$. The dissipative nature, introduced by the damping term, ensures that this search is directed towards a minimum, in full alignment with the variational objective of VQE.

\subsection{Comparison with Other Optimization Methods and Interpretation of Results}
The numerical experiments on both the hydrogen and lithium hydride molecules have highlighted the potential advantages of the velocity Verlet algorithm for VQE. Compared to conventional methods like L-BFGS-B, SLSQP, and COBYLA, the velocity Verlet optimizer demonstrated a capacity for achieving higher accuracy, particularly in the more challenging 12-qubit LiH system. This enhanced performance can be attributed to the inherent characteristics of the velocity Verlet method, namely its incorporation of "velocity" and "inertia" into the optimization process.

The inertia term allows the optimization trajectory to maintain momentum, which appears to be particularly beneficial in VQE's complex energy landscapes. As evidenced by the rapid initial energy reduction in the H$_2$ simulations (Figure~\ref{fig:1}), this momentum facilitates a more efficient exploration of the parameter space. For the more rugged landscape of LiH, this property may have enabled the optimizer to traverse shallow local minima or smooth out oscillations in steep valleys that could potentially trap optimizers relying solely on local gradient information. In contrast, methods like L-BFGS-B and SLSQP, which depend on instantaneous gradient and curvature information, can be more susceptible to being deflected by small landscape features, potentially leading to premature convergence in a suboptimal region.

However, the velocity Verlet method also presents a distinct set of trade-offs. The most significant is the computational cost associated with our current implementation. By requiring two gradient evaluations per iteration, the velocity Verlet method incurred the highest total number of energy evaluations for the LiH system. In the context of near-term quantum computing, where each energy evaluation is a costly resource, this is a critical consideration. Nonetheless, our results also suggest a nuanced choice for practitioners: for H$_2$, the velocity Verlet method was actually more efficient in terms of evaluations needed to reach chemical accuracy. For LiH, it was the only method to reach the lowest final energy, suggesting that a higher quantum cost might be a necessary price for achieving a desired accuracy on complex problems. 

\subsection{Hyperparameter Sensitivity and Tuning Strategies}
The performance of the velocity Verlet algorithm exhibits a significant sensitivity to the choice of its three hyperparameters: mass ($m$), time step ($\Delta t$), and damping coefficient ($\gamma$). This reliance on careful tuning represents a notable challenge compared to optimizers with fewer adjustable parameters. Each hyperparameter governs a key aspect of the optimization dynamics, and finding a synergistic balance among them is crucial for achieving robust and efficient convergence.

The mass parameter, $m$, directly controls the strength of the system's inertia. A larger mass causes the optimizer to retain the influence of past gradients for a longer duration, effectively increasing its momentum. This can be advantageous for promoting exploration and overcoming small energy barriers. However, excessive inertia, resulting from a mass that is too large, may cause the system to overshoot minima or to converge very slowly once a basin is found. Conversely, if the mass is too small, the method behaves more like standard gradient descent, losing its primary inertial advantage and becoming overly sensitive to local gradient fluctuations.

Similarly, the time step, $\Delta t$, which is analogous to the learning rate in other iterative methods, dictates the magnitude of parameter updates in each cycle. A larger $\Delta t$ can accelerate convergence by allowing for more substantial steps through the parameter space, but an excessively large value risks instability and divergence, particularly in regions of high landscape curvature.

Finally, the damping coefficient, $\gamma$, regulates the rate of kinetic "energy" dissipation. A value of $\gamma$ close to 1 corresponds to weak damping, preserving momentum across multiple iterations, which can be useful for traversing flat regions. In contrast, a value close to 0 signifies strong damping, which rapidly reduces the velocity, thereby enhancing stability. Insufficient damping may fail to suppress oscillations, especially in narrow, steep valleys, while excessive damping can prematurely halt exploration by diminishing the beneficial effects of inertia.

Achieving a careful balance among these competing effects is therefore essential. The optimal set of hyperparameters is inherently problem-dependent, as empirically demonstrated by the different mass parameters required for the H$_2$ and LiH systems in our study. While we selected these values through preliminary tuning, a key direction for future work is the development of adaptive strategies. Such methods could dynamically adjust these hyperparameters based on the local topology of the energy landscape—for instance, by reducing the time step or increasing damping upon detecting oscillations—thereby improving the algorithm's robustness and general applicability. 

\subsection{Potential Applications and Implications}
The velocity Verlet-enhanced VQE appears particularly promising for molecular calculations characterized by complex energy landscapes. This includes systems with many nearly-degenerate electronic states, molecules with flat potential energy surfaces where gradients are small, or large molecules where the parameter space is vast and contains numerous local minima. By improving the likelihood of finding a more accurate ground state energy, this approach could contribute to more reliable property predictions for new materials and more accurate calculations of drug-target binding energies in computational drug discovery.

Looking forward, several avenues for improvement exist. A key priority is to reduce the per-iteration evaluation cost. This can be readily achieved by implementing a Leapfrog variant of the algorithm, which is mathematically similar but requires only a single gradient evaluation per step. This would immediately halve the gradient-related quantum cost, making the method far more competitive in terms of total evaluations. Furthermore, combining the velocity Verlet optimizer with more advanced, hardware-aware gradient estimation techniques beyond the parameter-shift rule could yield additional savings. Finally, the true test of this method will be its application to larger molecular systems and its performance on real, noisy quantum hardware, where the interplay between algorithmic dynamics and physical device noise will be a critical factor to investigate. 

\section{Conclusion}

This study investigated the application of the velocity Verlet algorithm to optimize parameters in the VQE. By introducing "velocity" and "inertia" concepts, the velocity Verlet method offers potential advantages over standard gradient descent and other common optimizers. Numerical simulations on the H$_2$ molecule demonstrated that Velocity Verlet can achieve faster convergence and lower final energy within a fixed iteration limit.

While promising, these results are preliminary. Further research is needed to evaluate the performance of Velocity Verlet on larger, more complex molecules, and under realistic noisy conditions. The sensitivity of Velocity Verlet to its hyperparameters (mass, time step, damping) also requires careful consideration and systematic tuning strategies. Exploring more efficient gradient estimation techniques and different ansatz architectures is also crucial.

Despite these limitations, the results suggest that the velocity Verlet algorithm is a promising approach for enhancing VQE's efficiency and accuracy. Future work combining Velocity Verlet with other optimization techniques, and ultimately, testing on real quantum hardware, will be critical to realize its full potential for applications in quantum chemistry and materials science.

\section*{Declarations}

\noindent \textbf{Data availability}\\
All parameters used in the simulations are detailed in the Methods section and/or are available within the provided code.

\noindent \textbf{Code availability}\\
All program codes are available free of charge on GitHub at \url{https://github.com/G0wOz1PV/VQE}. All supported systems are equipped with Python 3.7 or higher.

\noindent \textbf{Competing interests}\\
The authors have no relevant financial or non-financial interests to disclose.

\bibliography{main}

\end{document}